\documentclass[preprint,pra,showpacs,superscriptaddress]{revtex4-1}
\usepackage{amssymb}
\usepackage{mathrsfs}
\usepackage{lipsum}
\usepackage{graphicx}
\usepackage[caption=false]{subfig}
\usepackage{mathtools}
\usepackage{epstopdf}
\usepackage{xcolor}
\usepackage[colorlinks=true]{hyperref}
\DeclareGraphicsExtensions{.pdf,.jpg,.png,.eps}
\usepackage[toc,page,title,titletoc,header]{appendix}
\begin{document}

\title{Cavity mediated effective interaction between phonon and magnon}

\author{Yong-Pan Gao}
\address{School of Science and the State Key Laboratory of Information Photonics and Optical Communications, Beijing University of Posts and Telecommunications, Beijing 100876, China}
\author{Cong Cao}
\address{School of Ethnic Minority Education, Beijing University of Posts and Telecommunications, Beijing 100876, China}
\author{Tie-Jun Wang}
\address{School of Science and the State Key Laboratory of Information Photonics and Optical Communications, Beijing University of Posts and Telecommunications, Beijing 100876, China}
\author{Yong Zhang}
\address{School of Science and the State Key Laboratory of Information Photonics and Optical Communications, Beijing University of Posts and Telecommunications, Beijing 100876, China}
\author{Chuan Wang}
\address{School of Science and the State Key Laboratory of Information Photonics and Optical Communications, Beijing University of Posts and Telecommunications, Beijing 100876, China}
\vspace{10pt}

\begin{abstract}
Optomagnonics and optomechanics have various applications ranging from tunable light sources to optical manipulation for quantum information science. Here, we propose a hybrid system with the interaction between phonon and magnon which could be tuned by the electromagnetic field based on the radiation pressure and magneto-optical effect. The self energy of the magnon and phonon induced by the electromagnetic field and the influence of the thermal noise are studied. Moreover, the topological features of the hybrid system is illustrated with resort to the dynamical encircling with the exceptional points, and the chiral characteristics under these encirclements are found.
\end{abstract}
\maketitle
%
%
%
%
%

\section{Introduction}
Optomechanics \cite{Bowen-Milburn,RevModPhys.86.1391,Aspelmeyer,Kippenberg:07,peng2014and,anetsberger2009near,Kippenberg1172,hill2012coherent,PhysRevLett.110.153606,PhysRevLett.98.150802,thompson2008strong} and optomagnonics \cite{PhysRevLett.113.156401,PhysRevB.94.060405,PhysRevLett.117.123605,PhysRevA.92.063845,PhysRevB.93.144420,PhysRevLett.116.223601,PhysRevLett.117.133602,PhysRevA.94.033821,klos2014photonic,chumak2015magnon,Tabuchi405,PhysRevLett.111.127003} are both hybrid optical interaction which could be achieved on the microresonator system. Efficient and low-noise optomechanics has many applications
in classical and quantum photonics, such as squeezed light generation \cite{brooks2012non,safavi2013squeezed,PhysRevX.3.031012,PhysRevA.83.033820,PhysRevA.82.021806}, mass and force sensing \cite{PhysRevLett.97.133601,PhysRevLett.105.123601,teufel2009nanomechanical,PhysRevLett.113.020405}, optical manipulation \cite{PhysRevLett.112.013602,Weis1520,monifi2016optomechanically,safavi2011electromagnetically,jing2015optomechanically}, and so on. Moreover, optomechanics also provides us a platform to study the basic tasks of quantum information processing \cite{PhysRevLett.109.013603,Ying-Dan,PhysRevLett.111.083601,PhysRevLett.109.063601}. On the other hand, the magnon is the quantized magnetization excitation in magnetic materials \cite{Serga,PhysRevLett.110.147206,lenk2011building} and optomagnonics describes the interaction between photon and the magnon through the magnetic dipole interaction. It have been emerged recently to achieving the coherent information processing based on the optomagnonics \cite{Cowburn1466,Imre205,PhysRevLett.103.043601}. Due to the characters of long lifetime \cite{PhysRevLett.103.043601,PhysRevLett.74.1867,zhang2015magnon}, the magnon is a good candidate for information storage.

The first description of the optomechanical effect can be traced back to 1619, Johannes Kepler put forward the concept of radiation pressure to explain the observation that a tail of a comet always points away from the Sun. With the development of the microfabrication technology, various schemes based on the microresonator optomechanical system have been theoretically proposed and experimentally achieved with the increment of the cavity quality factor and the decrement of the scale  \cite{armani2003ultra,vahala2003optical,gorodetsky1999optical}. The most famous study of magneto-optical (MO) effects \cite{Magneto-Optics} is the Faraday effect \cite{PhysRev.137.A448,inoue1997theoretical} which has been utilized in discrete optical device applications \cite{RevModPhys.82.2731,kimel2005ultrafast}. Based on the MO effect, magnons can be generated and probed under the process of optical pumping \cite{Demokritov2001441,PhysRev.166.514}. Most recently, the optical-magnetic interaction has been experimentally achieved with the magnetic insulator yttrium iron garnet(YIG) doped micro-sphere cavity in the $c$ band \cite{PhysRevLett.117.123605}. In this experiment, the magnon-phonon coupling strength can reach $100$ kHz level in the cavity with the quality-factor (Q-factor) of $10^6$.


In this paper, we consider a high-quality YIG doped micro-sphere cavity with both optomechaical and optomagnetical properties.  Here we assume the electromagnetic field works in the whispering-gallery mode(WGM), and the transverse-magnetic (TM) and transverse-electric (TE) modes have slightly frequency detuning as there is a slight difference in their effective refractive index \cite{GORODETSKY1994133}. With the independently tunable frequency, the TE and TM modes play different roles in our system. We find that the electromagnetic field can add the self-energy component onto the magnon and phonon, and even bridge the interaction between phonon and magnon. Also the chiral topological characteristics are studied.


This article is structured as follows: In section \ref{sec:system hamitonian}, we give the linearized Hamiltonian and the coupling equations in the frequency domain. Based on these equations, we obtain the self-energy and the medicated interaction in section \ref{sec:selfenergy}. In section \ref{sec:monitor}, we describe a method to manipulate the thermal environment of the magnon and phonon. In section \ref{sec:energy}, the eigen-energy of the magnon-phonon hybrid system are given, and we find the Riemann surface style energy scheme under the input field medicating. To further study the topological feature, we show the dynamic adiabatic evolution of this system in section \ref{sec:toplogical}.

\section{System Hamiltonian}
\label{sec:system hamitonian}

\begin{figure}[!htbp]
\centerline{\includegraphics[width=0.4\textwidth]{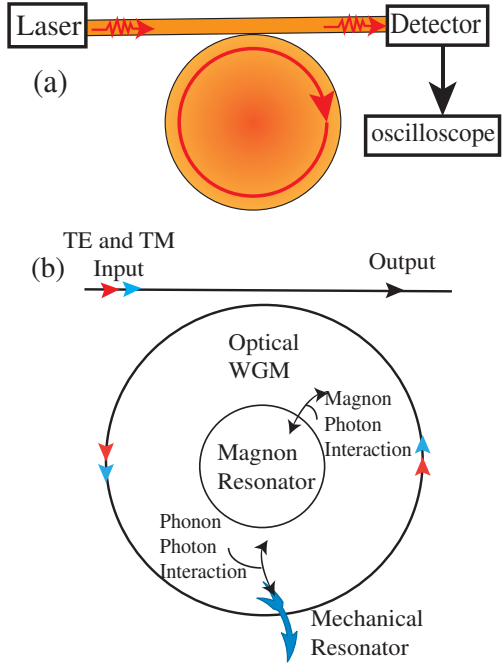}}
\caption{(color online)The structure of the studied system in this article. Fig.(a) shows the consist objects of this system, a glass ball doped with YIG have the mechanical mode. A fiber taper coupled the pumped field with the magnon-phonton-phonon system. In Fig.(b) the interaction scheme is shown. Through the phonon-photon and the magnon-photon interaction. The Magnon and phonon system were bridged. The red and blue arrows show represent the TE and TM pumping respectively.}
\label{structure}
\end{figure}

 As shown in Fig.\ref{structure}, the system under consideration is a YIG doped micro-sphere cavity. With the YIG doping, the electromagnetic field interacts with the magnetic resonator which provides phonton-magnon interaction. The glass sphere has its intrinsic mechanical frequency, which indicates that the intracavity field can drive the phonon-photon interaction.
The effective Hamiltonian under the non-depletion approximation and linearization have the form (The detail deduction can be found in \ref{app:A} )
\begin{equation}
H=\omega_{b}\hat{b}^{\dag}\hat{b}+\omega_{m}\hat{m}^{\dag}\hat{m}+\omega_{r}\hat{r}^{\dag}\hat{r}+G_{a}\hat{b}^{\dag}\hat{m}+G_{a}^{*}\hat{b} \hat{m}^{\dag}+G_{b}(\hat{b}^{\dag}+\hat{b})\hat{z},
\label{Hamiltan}
\end{equation}
where $G_{a}=g_{m}<a>=g_{m}\sqrt{2\kappa_{a,1}}a_{in}/(-i(\omega_{L}-\omega_{a})+\kappa_{a})$ is the effective magnon-photon coupling strength, while $G_{b}=g_{b}<b>=g_{b}\sqrt{2\kappa_{b,1}}b_{in}/(-i(\omega_{pu}-\omega_{b})+\kappa_{b})$ is the effective phonon-photon coupling strength.  $\omega_{a}$($\omega_{b}$) is the frequency of the TM(TE) mode, $\omega_{m}$ is the magnetic frequency and $\omega_{r}$ is the frequency of the phonon. The $j^{\dag}$($j$) $(j=a,b,m,r)$ is the creation (annihilation) operator of the related mode (TM mode photon, TE mode photon, magnon and phonon). $g_{m}$ describes the interaction strength between the photon and magnon, and $g_{b}$ describes the interaction strength between photon and phonon. $\hat{z}=\hat{r}^{\dag}+\hat{r}$ is the displacement operator of the phonon. Based on the Hamiltonian, we can get the coupling equation on the frequency domain as (\ref{App;B})
\begin{subequations}
\begin{align}
\chi_{b}^{-1}[\omega]b[\omega]&=-iG_{b}z[\omega]-iG_{a}m[\omega],\label{fourierb}\\
\chi_{r}^{-1}[\omega]r[\omega]&=-i(G_{b}^{*}b[\omega]+G_{b}b^{*}[-\omega])+\sqrt{\gamma_{r}}\eta_{r}[\omega],\label{fourierr}\\
\chi_{m}^{-1}[\omega]m[\omega]&=-iG_{a}b[\omega]+\sqrt{\gamma_{m}}\eta_{m}[\omega].
\label{fourierm}
\end{align}
\label{fourier}
\end{subequations}
The $\chi[\omega]=(\gamma/2-i(\omega-\omega_{j})),j=b,r,m$ is the intrinsic susceptibility of its corresponding degree of freedom.

\section{The optical modulation of phonon and magnon interaction}

\subsection{The mediated self-energy and interaction strength}
\label{sec:selfenergy}
The case of optically medicated mechanical resonator has been studied previously \cite{PhysRevLett.112.013602}. In order to investigate the self-energy of  the magnon and the mediated interaction, we eliminate the electromagnetic part $b[\omega]$ of the system and solve the self-energy part of the system. And we neglect the effective interaction terms between the phonon and magneton,
\begin{subequations}
\begin{align}
(\chi_{m}^{-1}[\omega]+i\Sigma_{mm}[\omega])m[\omega]&=\sqrt{\gamma_{m}}\eta_{m}[\omega],\\
(\chi_{r}^{-1}[\omega]+i\Sigma_{rr}[\omega])z[\omega]&=\sqrt{\gamma_{r}}\eta_{r}[\omega],
\end{align}
\label{selfenergy}
\end{subequations}
where $\Sigma_{rr}[\omega]=-i |G_b|^2(\chi_{b}[\omega]-\chi_{b}^{*}[-\omega])$ denotes the optomechanical self-energy. Similarly, $\Sigma_{mm}[\omega]=-i G_{a}^2\chi_{b}[\omega]$ is defined as the self-energy of the magnon. This self-energy represents the contribution  of the optomagnonics to the magnonical resonance frequency $\delta\omega_{i}=\mathrm{Re}(\Sigma_{mm}[\omega_{i}])$ and damping $\delta\gamma_{i}=\mathrm{Im}(\Sigma_{mm}[\omega_{i}]), i=m,r$. Here  $G_{a}$ is mediated by the TM pumping field and $\chi_{c}$ can be controlled by the detuning $\Delta_{b}$ of the TE optical pump field.

To study the relationship between the frequency detuning and self-energy, we assume that the $TE$ mode is pumped with the frequency $\omega_L =\omega_b$ and the magnon  resonant with the phonon ($\omega_{r}=\omega_{m}$). By selecting the proper Q-factor, we have the relation $\gamma_{m}=\gamma_{r}=\kappa_{a}=\kappa_{b}=\gamma$. Suppose both the TE and TM modes are under the critical coupling condition as $\kappa_{i,1}=\kappa_{i}/2 (i=a,b)$ with the pumping field. Besides, we take $a_{in}g_{m}=b_{in}g_{b}$ to ensure $G_a$ is comparable with $ G_b$. Here we set $\omega_{r}=0.4\gamma$, then the mechanical mode is working on the unresolved-sideband regime.
\begin{figure}[!htbp]
\centerline{\includegraphics[width=0.8\textwidth]{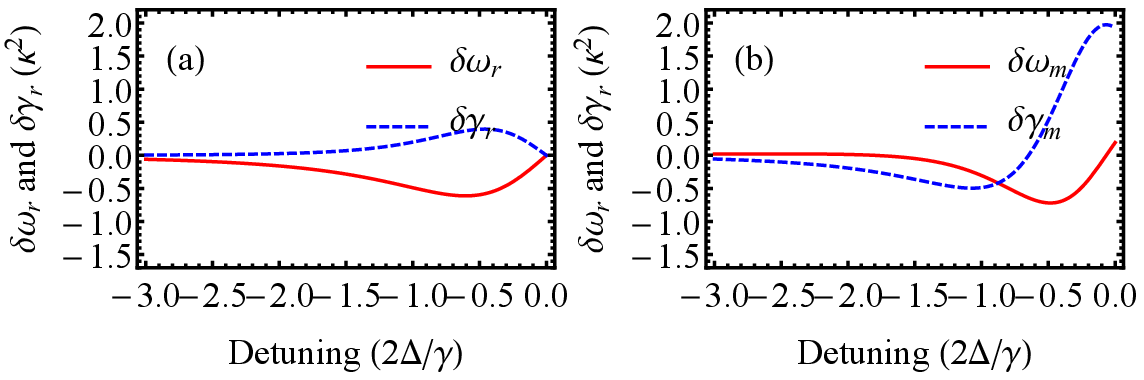}}
\centerline{\includegraphics*[width=0.8\textwidth]{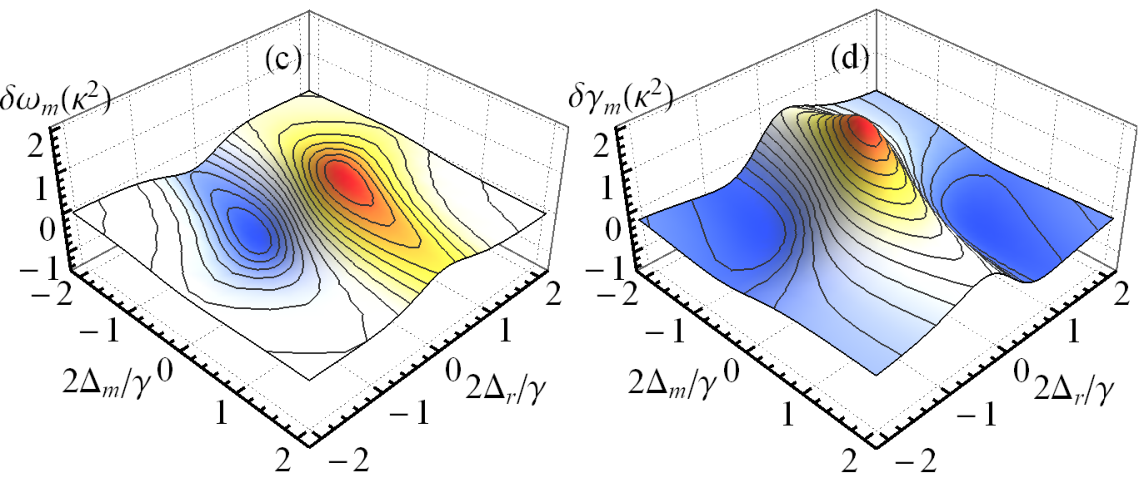}}
\caption{The strength of the self energy as a function of detune. In (a) and (b) we plot the frequency and the damping shift under different monochromatic driving, while in (c) and (d) we respectively plot the frequency and the damping shift of the magnon when the system is driven with different frequency on TM and TE mode. Here the detuning-$\Delta$ is plot in the step of decay $\gamma$. The amplitude of the frequency and the damping shift have a unit of $(g_{i}a_{i}\kappa)^2$. Here (a) shows the frequency and damping shift of the mechanical resonator. As symbol in the figure, the red solid line corresponding toe the frequency shift while the blue dash line shows the damping modulation. The figure (b) shows the same subjects of the magnon's.}
\label{inout}
\end{figure}
To intuitively show the influence of the pump field frequency on the self-energy, we plot the medicated frequency and damping in Fig.\ref{inout}. Compared with the self energy of the mechanical resonator shown in Fig.\ref{inout}(a) \cite{PhysRevLett.112.013602}, the self energy of the magnon is different. In Fig.\ref{inout}(b), we plot the monochrome pumped condition (The TM pump frequency is equal to TE pump frequency). It shows the difference of self energy between the magnon and mechanical resonator, both the medicated frequency and the damping of the magnon can be positive and negative valued. In general, the positive and negative frequency detune corresponds to the blue and red shift of the resonator, while for the damping term, the positive and negative mean the resonator work in the gain or dissipative region.  So it is shift and damping features tunable for magnon but not for the mechanical resonator. In order to further study the characteristics of the self-energy of the magnon. We plot the 3D-graph to show the influence of the frequency of the TE and TM field. Fig.\ref{inout}(c) shows the frequency shift of the magnon, while in Fig.\ref{inout}(d) we show the dissipation's variation under different TE and TM pump detuning. As shown in these two figures, the action of the frequency shift and damping change are not synchronization. In this sense, we can independently control the frequency and damping.

The self-energy term shows the interaction between the resonators and the optical field, while the interaction terms provides the interface of the magnon and phonon. To study its influence, we rewrite the Eq.\ref{selfenergy} with the consideration of the medicated interaction
\begin{subequations}
\begin{align}
(\chi_{m}^{-1}[\omega]&+i\Sigma_{mm}[\omega])m[\omega]=\sqrt{\gamma_{m}}\eta_{m}[\omega]-i\Sigma_{mr}[\omega]z[\omega],\\
(\chi_{r}^{-1}[\omega]&+i\Sigma_{rr}[\omega])z[\omega]=\sqrt{\gamma_{r}}\eta_{r}[\omega]-i\Sigma_{rm}[\omega]m[\omega]+i\Sigma^{*}_{rm}[-\omega]m^{*}[-\omega]).
\end{align}
\label{effective}
\end{subequations}
Here $\Sigma_{mr}[\omega]=-i G_{a} G_{b} \chi_{b}[\omega]$ and  $\Sigma_{rm}[\omega]=-i G_{a} G^{*}_{b} \chi_{b}[\omega]$ are the medicated interaction strength. The coupling has different coupling dissipation. Meanwhile, the $G_a$ and $G_b$ can be medicated by the TE and TM pumping frequency. Then we can control the interaction strength by modulate the pump field. In order to simplify the results, we set $\omega_{r}=\omega_{m}$, $\gamma_{m}=\gamma_{r}=\kappa_{a}=\kappa_{b}=\gamma$ and $\kappa_{j,1}=\kappa_{j}/2 (j=a,b)$.
\begin{figure}[!htbp]
\centerline{\includegraphics*[width=0.5\textwidth]{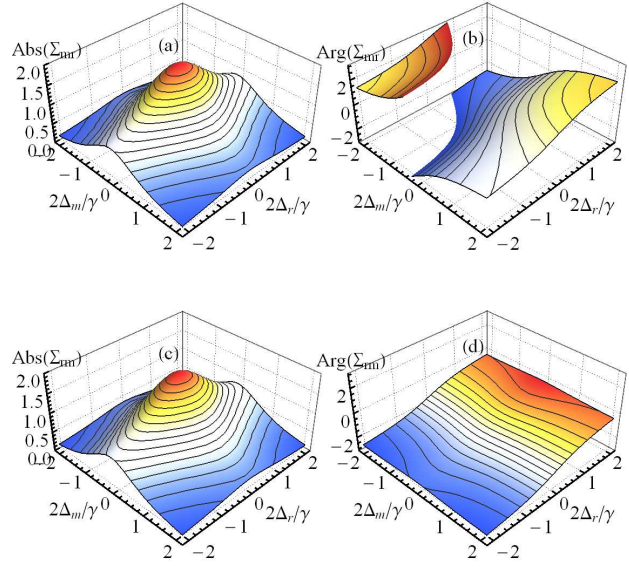}}
\caption{(color online)The real and image part of the medicated coupling strength. We show the $\Sigma_{mr}$ in (a) and (b), while $\Sigma_{rm}$ in (c) and (d). Here (a) and (c) shows the real part, while (b) and (d) shows the image part. Here the detuning-$\Delta$ is plot in the step of decay $\gamma$. The amplitude of the frequency and the damping shift have a unit of $(g_{i}a_{i}\kappa)^2$. The subscript $ij$ means the effect of $i$ to $j$ system.}
\label{interaction}
\end{figure}
In Fig.\ref{interaction} we respectively plot the strength and phase part of the interaction strength of mechanical resonator to magnon and magnon to mechanical resonator. In the figure, Fig.\ref{interaction}(a) and Fig.\ref{interaction}(b) show the strength and phase of $\Sigma_{mr}[\omega_i]$ respectively, while the strength and phase of $\Sigma_{rm}[\omega_i]$ in exhibited in Fig.\ref{interaction}(c) and Fig.\ref{interaction}(d). It can be easily find that these two coupling strength have the same strength. We know that absolute phase of the interaction will not influence the features of the system. But as shown in  Fig.\ref{interaction}(b) and Fig.\ref{interaction}(d). The coupling phase will be different in most region of this system. Then there will be a relative phase. From Eq.\ref{effective}, we can also elicit that there is a phase difference origin from the $G_{b}$ and $G_{b}^{*}$ term. This relative phase gives us one more freedom to control the system. Even more the generation of this  relative phase is also a novel feature of this hybrid system.

\subsection{The transverse electric field transmission}
\label{sec:monitor}

To further describe the magnon-phonon system using the electromagnetic field, we first give the expression of the electromagnetic field of the TE mode as
\begin{subequations}
\begin{align}
b[\omega]=&\frac{X[\omega]Y[\omega]X^{*}[-\omega]+X[\omega]}{1-X[\omega]Y[\omega]X^{*}[-\omega]Y^{*}[-\omega]}Z[\omega],\\
\begin{split}
X[\omega]=&(\chi_{b}^{-1}[\omega]+G_{b}^{*}G_{b}(\chi_{r}[\omega]-\chi_{r}^{*}[-\omega])\\&+G_{a}^{2}\chi_{m}[\omega])^{-1},\end{split}\\
Y[\omega]=& G_{b}^{2}(\chi_{r}[\omega]-\chi_{r}^{*}[-\omega]),\\
Z[\omega]=&-iG_{b} (\chi_{r}[\omega]+\chi_{r}^{*}[-\omega])\sqrt{\gamma_{r}}\eta_{r}
-iG_{a}\chi_{m}[\omega]\sqrt{\gamma_{m}}\eta_{m}.
\end{align}
\label{electromagnetic}
\end{subequations}
In these equations, the effective photon-magnon input term $Z[\omega]$ is affected by the environment of the magnentic and mechanical resonator which is described by the expression of the electromagnetic field $b[\omega]$ as shown in Eq.\ref{electromagnetic}. Here we take the thermal noise environment as an example to study the feature of this system. The magnetic and mechanical mode of the resonator has the quality factor $Q_r=Q_m\sim1\times10^3$. Consider the  stochastic and incoherent feature, we assume that
\begin{equation}
S_{Z}[\omega]=\delta_{r,m}Z^{*}[\omega_1]Z[\omega].
\end{equation}
  The thermal noise spectrum is $k_{B}T/\omega$ which could be set as a constant as the change of the frequency ($|\omega_{r}-\omega_{m}|$) is neglectable when compared with the frequency we studied ($\omega_{r},\omega_{m}$). The thermal noise is set as the unit value as the effect of the noise exists on the whole system. Here the two terms $a_{in}g_{m}$ and $b_{in}g_{r}$ are chosen as a whole part. Consider that $a_{in}g_{m}\sim b_{in}g_{r}\sim 1$ THz and $g_{m}\sim100Hz$ \cite{PhysRevLett.117.123605}, $g_{m}\sim5000Hz$ \cite{PhysRevLett.112.013602} and the wavelength is around 1550nm, we can calculate that the TM mode pumping power $P_{m}\sim20$ mW while the TE pumping power is  $P_{r}\sim7.9\mu$ W. And the damping rate of this system are set as $\kappa_{a}=\kappa_{b}=\kappa_{m}=\kappa_{r}=20$MHz. The magnon and photon have the frequency of $\omega_{m}=1Gz-150$MHz and $\omega_{r}=1$Gz$+150$MHz.

\begin{figure*}[!htbp]
\includegraphics[width=1.\textwidth]{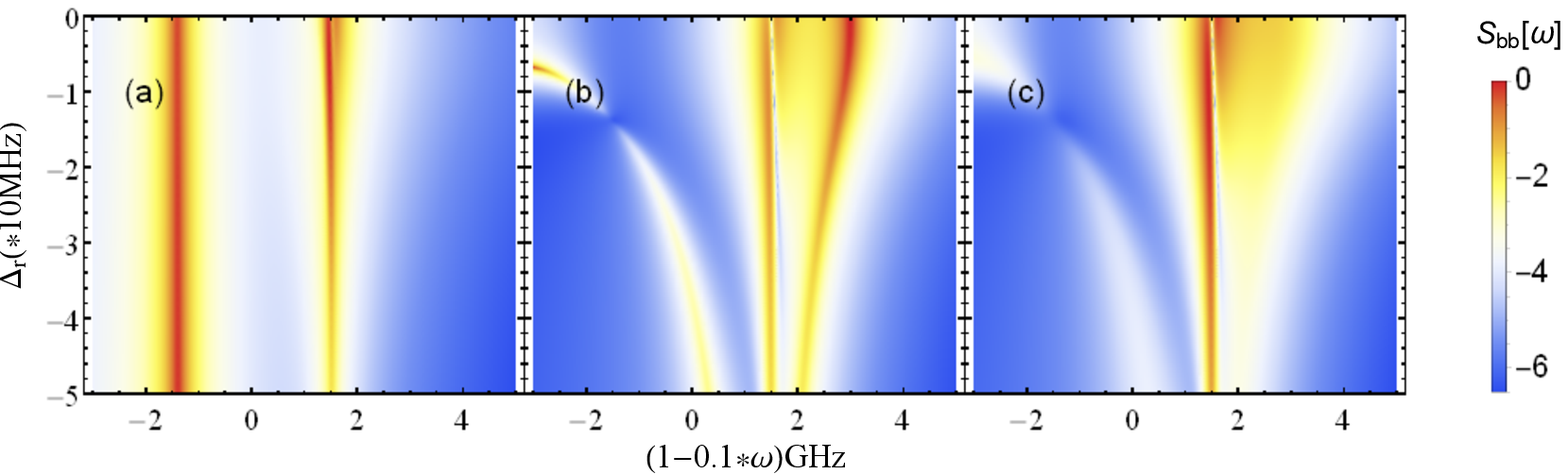}
\includegraphics*[width=1.\textwidth]{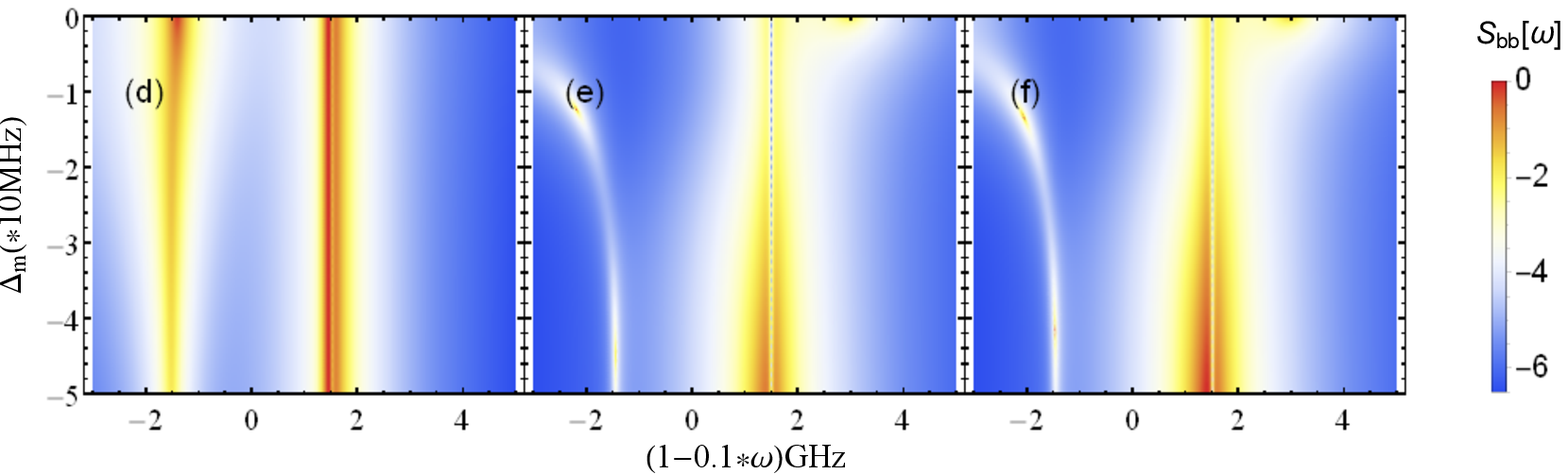}
\caption{ Power spectral density (arbitrary units) of the heterodyne signal. (a)(b) (c)TE and (d)(e)(f)TM as a function of measurement frequency (horizontal axis) and detuning between the incident laser and the cavity resonance (vertical axis). The date in (a) and (d) shows the low effective input power $a_{in}g_{m}=b_{in}g_{r}=0.6$ THz, this power is $a_{in}g_{m}=b_{in}g_{r}=3.6$ THz in figure (b),(c),(e)and (f). In figure (a) and (b) The TM detune is set as $\Delta_{m}=0Hz$ while in (c)$\Delta_{m}=-3MHz$. In figure (d) and (e) The TE detune is set as $\Delta_{m}=0Hz$ while in (f)$\Delta_{m}=-3MHz$.  }
\label{optical}
\end{figure*}
As the intrinsic susceptibility of the system is sensitive to the frequency, the thermal noise will drive this system effectively on the resonant condition. 
In this structure, both the pump frequency and the pump power of the electromagnetic field can be modulated, and even the TE and TM field.
We further study the system by using the optical spectrum under the influence of the thermal noise in Fig.\ref{optical}. In Fig.\ref{optical}(a), we plot the effective pumping $a_{in}g_{m}=b_{in}g_{r}=0.6$THz, while we set the detuning of the TM pumping field is zero($\Delta_{m}=0$). The thermal noise can be observed in the frequency domain of the TE mode. In this figure, the maximal value of the electromagnetic field corresponds to the condition that the thermal noise resonant with the phonon-magnon system, then the effective frequency of the phonon-magnon hybrid resonator could be readout. While in Fig.\ref{optical}(d), we show the influence of TM detuning on the spectrum.

By increasing the pumping power, the optically-medicated effect becomes obvious. The electromagnetic field works as an optical spring in this condition which  connects the phonon and magnon system by producing an effective coupling between them. In order to discuss this strong pumping region, we enlarge this effective driving strength to $a_{in}g_{m}=b_{in}g_{r}=3.6\times10^{12}$Hz. Fig.\ref{optical}(b)(c) show the influence of the TE pump detuning. In Fig.\ref{optical}(b) the TM detuning is set as $\Delta_{m}=0Hz$, while it is $\Delta_{m}=-3$MHz in Fig.\ref{optical}(c). Three bands could be observed in this system, the magnon band, the phonon band and the optical spring band. The left band corresponds to the magnon mode and there is an obvious frequency shift in the magnon mode. The magnon mode disappears when the detune is bigger than $-10$MHz. It means that the magnon mode becomes a dark mode. The interaction between the magnon mode and the optical field disappears while it interferes with the optical and mechanical hybrid interaction. In Fig.\ref{optical}(c), we can find that the third band obviously changes with a slight detuning of TM mode. We can conclude that the right band can be connected with the optical spring effect.

The influence of TM detuning is shown in Fig.\ref{optical}(e)(f). We can find the third band only can be found in the zero detuning region. In these two figures, the magnon mode is weaker than the phonon mode, and it comes from the relative coupling phase of the magnon-photon to phonon-photon. And we can find  both in (e) and (f), the dark magnon state appears when the pumping detune is bigger than $-10$MHz. And when the TM detuning is large enough, the effective frequency shift is not so obvious. It can be explained as when the detune is strong, the magnon is decoupled from this system, then the mognon mode could not be found in the optical field. Different from Fig.\ref{optical}(e) with $\Delta_{r}=0$, Fig.\ref{optical} (f) is plotted under the detuning $\Delta_{r}=-3$MHz. We can find that this two optical spectrum are identical. The TE detuning only influence the absolute intensity of the field in the plotted region.

\section{The topological features of the hybrid phonon-magnon system}

\subsection{Topological energy with exceptional points}
\label{sec:energy}

\begin{figure*}[!htbp]
\centerline{\includegraphics*[width=0.9\textwidth]{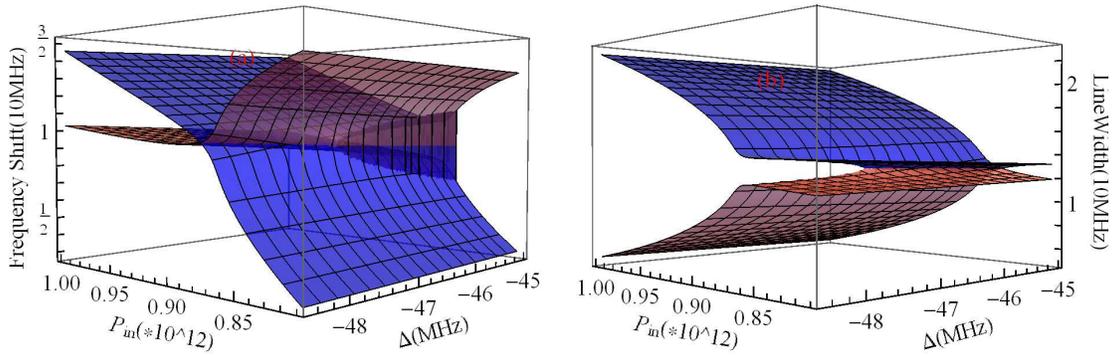}}
\caption{(color online) Then energy scheme of this hybrid system. Figure (a) show the real part (b) is the image part.Here the $P_{in}=a_{in}g_{m}=b_{in}g_{r}$ is the effective input strength, while the $\Delta$ is the frequency detune of the TE mode. The TM mode detune is set as $-3MHz$. We use $\Delta$ to symbol $Delta_{b}$-the TE detune. Here the cavity, magnon and phonon  dissipation is $20MHz$. The magnon has the frequency of $1GHz-16MHz$, while it is $1GHz+16MHz$ for the phonon. }
\label{eigenvalue}
\end{figure*}

We consider the system combines the magnon and phonon as a closed system which include an optical pumping. For simplicity, we study the mechanical mode in the phonon mode but not displacement momentum operator. The reduced Hamiltonian can be write as (The detailed derivation can be find in  \ref{App;C} ),
 \begin{equation}
 H=\left\{\begin{matrix} \omega_{r}-i\frac{ \kappa_{r}}{2} & 0  \\ 0 & \omega_{m}-i \frac{ \kappa_{m}}{2} \end{matrix}\right\}\\
-i\left\{\begin{matrix} |G_{b}|^{2}[\chi_{b}[\omega_{m}]-\chi_{b}^{*}[-\omega_{m}]] & G_{a}^{*}G_{b}\chi_{b}[\omega_{m}]  \\ G_{a}G_{b}\chi_{b}[\omega_{m}] & G_{a}^{2}\chi_{b}[\omega_{m}] \end{matrix}\right\}.
 \label{hamilton}
 \end{equation}
With the exact diagonalization of the Hamiltonian, two eigenenergies are shown as a function of pumping power and detuning of the TE field in Fig\ref{eigenvalue}. This system will possess the exceptional points(EPs) if the system which is tuned at $16|G_a|^{2}G_{b}^{2}\chi_{b}^{2}+(2|G_{b}|^{2}(\chi_{b}^{*}-\chi_{b})+2G_{a}^{2}G_{b}+\kappa_{m}-\kappa_{r}+2i(\omega_{m}-\omega_{r}))^{2}=0$. To achieve the equality condition, both the real and image part should be controlled. Fortunately, $G_{a}$ and $G_{b}$ are both pumping frequency and strength dependent, while $\chi_{b}$ can also be controlled by the TE pumping frequency. To simplify this study, we set the TM detuning equal to the TE detuning, even more, the effective pumping rate are set as equal values.

In Fig.\ref{eigenvalue}, we show the real and imaginary part of the eigenenergy of the medicated Hamiltonian in Eq.\ref{hamilton}. And we take the parameters shown in last section, except that the magnon frequency is $984$MHz, while the phonon has frequency $1016$MHz. Even more, we fix the value of the TM detuning as $\Delta_{a}=-3$MHz. The input field could be written as the effective form $P_{in}=a_{in}\times g_{m}=b_{in}\times g_{r}$. For the eigenfrequency, we plot its shift from $1$GHz. Here the EPs are obviously in this figure. The two eigenenergy surface of the two super-mode form a Riemann surface, this gives the system a topological feature when the optical pumping encircling with differer direction of the surface. As shown in Fig.\ref{eigenvalue}(b), we find the dissipation can be negatively valued. This means the optical field can work as a gain medium, while the gain energy comes from the pumping field.

\subsection{Topological encircling and state evaluation}
\label{sec:toplogical}
In order to further study the topological features of this system, we encircling this system under different directions on the Riemann surface. We make this system encircling in a close  path, and study the difference when the EPs are included and exclusive in this circle. Take Fig.\ref{eigenvalue} as a reference, we choose the encircling center point as $P_{in}=0.87$THz, $\Delta=-5.5$MHz and $P_{in}=0.87$THz, $\Delta=-4.5$MHz. Here the $P_{in}=a_{in}g_{m}=b_{in}g_{r}$ is the effective input power, while $\Delta$ denotes the frequency detuning of the TE mode. The TM mode detuning is set as $-3$MHz. Here we set the encircling radius as one unit, while one unit is 0.1THz (1MHz) for the $P_{in}$ ($\Delta$) axes. Consider the feature of the topological surface, we also need encircling this system in different directions.
\begin{figure*}[!htbp]
\includegraphics[width=0.9\textwidth]{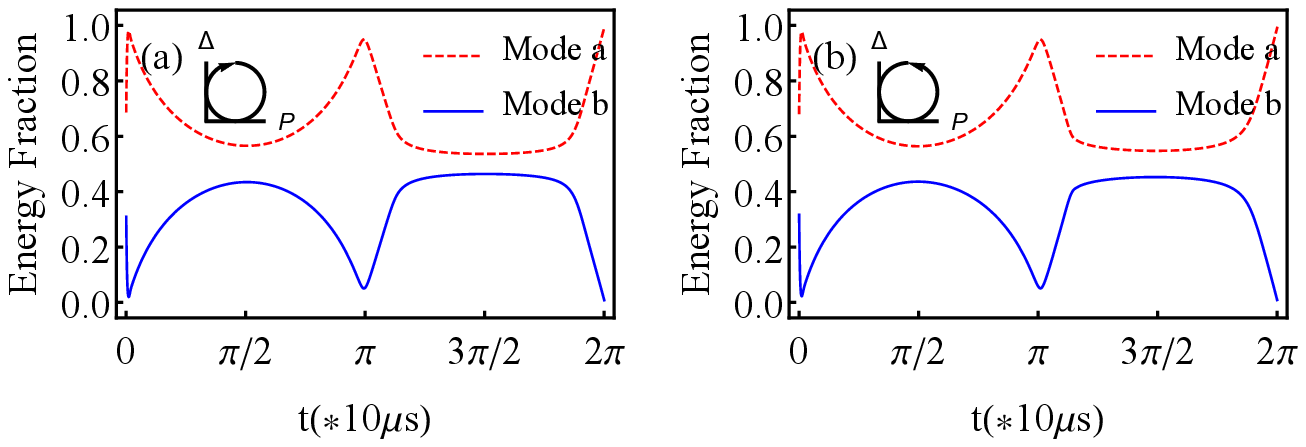}
\includegraphics*[width=0.9\textwidth]{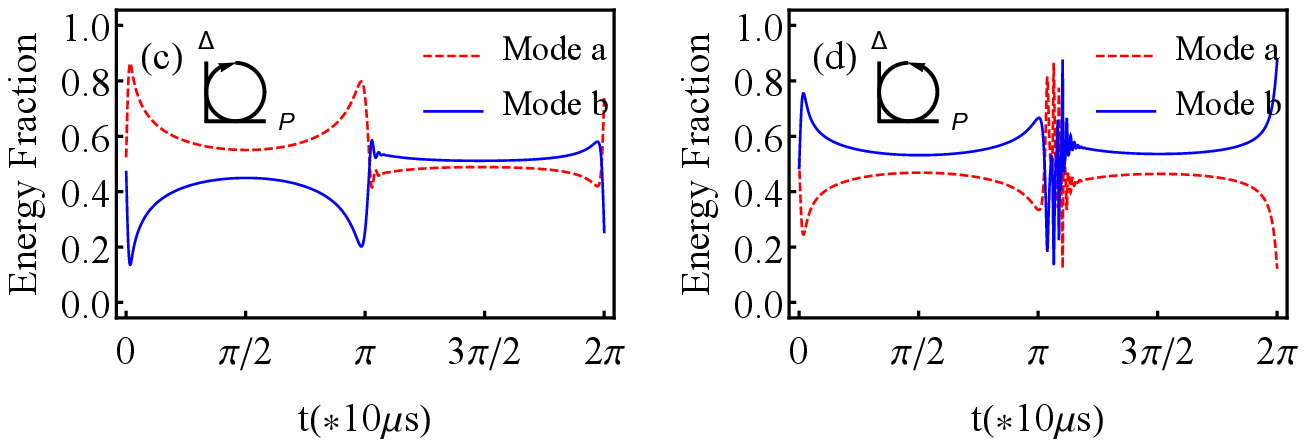}
\caption{Energy fraction under toplogical encircling. In this figure the effective input power encircling center $P_{in}=a_{in}g_{m}=b_{in}g_{r}=0.87$THz. We use $\Delta$ to symbol $\Delta_{b}$-the TE detune. Here the encircling center in $\Delta$-axes $\Delta=-5.5$MHz for figure (a) and (b), while it is $-4.5$MHz for (c) and (d). (a) and (b) encircling without EP point while  (a) and (b) including in it under this setting. The TM mode detune is set as $-3MHz$. The encircling radius is 1 unit. 1 unit is 0.1 THz for the $P_{in}$ axes while for the $\Delta$ axes its $1$MHz. Here the cavity, magnon and phonon  dissipation is $20$MHz. The magnon has the frequency of $1$GHz$-16$MHz, while it is $1$GHz$+16$MHz for the phonon. The encircling direction is as shown in the figure.}
\label{toplogical}
\end{figure*}

We plot the energy fraction of mode "a" and "b" of the studied system when encircling this system on the Riemann surface. Here mode "a" corresponds to the red surface of this system, while "b" corresponds to the blue part. In these figures, we always use the initial mode vectors to project the destiny matrix. Here in Fig.\ref{toplogical} (a) and (b), we plot the mode evaluation of the magnon-phonon super-mode when the system encircling with the direction shown in the figure. Also the red line corresponds to mode "a" while the blue line denotes mode "b". Here Fig.\ref{toplogical}(b) and (d) have an additional phase $\pi$ to make the mode evaluation synchronize with Fig.\ref{toplogical}(b). Here this system is encircling with the period of $T=10$ms to make sure the system  adiabatic evaluation for this system has a frequency of GHz.

Here we analyze the phenomenon shown in Fig.\ref{toplogical}(a) and (b). In these figures, the encircling region does not include the EP points. We can find the mode evolution almost in the same way. With the adiabatic evaluation, this system shows similar features when encircling in different direction. The tiny difference comes the system stays in different Riemann surface. Different from the above condition, when we include the EP points in the encircling, the evolution of this system will be quite different. As shown in Fig.(c)(d), the difference is that the system will experience an acute oscillation even this system is evolute in a slow way. After the oscillation, there is a crossover between the two energy levels. It means that there is an energy level variation during the evolution. The EP points are work as phase change points as the system feels a sudden energy change when encircling. Moreover, the system is different when it evaluating in different direction. The main differences of the oscillation are the strength and lasting time. It can be attribute to the EP point with chirality when encircling as shown in Fig.\ref{eigenvalue}.

\section{Summary}

In summary, we have theoretically investigated the electromagnetic cavity-mediated  phonon-magnon interaction system and elaborated how to connect the phonon and magnon mode through the optical medication. As a medium, the electromagnetic field will bring the self-energy and medicate the interaction between the magnon and phonon. When take the thermal noise as an example to study the influence of the environment, we find that the thermal field influences the output of the electromagnetic field. Even more, the magnon will become a dark state when the detune were small. Base on the coupling equations,  we investigate the topological features of the hybrid system with the Hamiltonian. When encircling with the EPs on the Riemann energy-surface, the energy level projection shows different vibration behaviour which indicate the chirality of the system.

\section{Acknowledgements}
The authors gratefully acknowledge the support from the National Natural Science Foundation of China through Grants Nos. 61622103, 61471050 and 61671083; the Ministry of Science and Technology of the People's Republic of China (MOST) (2016YFA0301304); the Fok Ying-Tong Education Foundation for Young Teachers in the Higher Education Institutions of China (Grant No. 151063); the Open Research Fund Program of the State Key Laboratory of Low-Dimensional Quantum Physics, Tsinghua University Grant No. KF201610. Also we would like to thank Dr. Qi-Kai He for helpful discussions.

\appendix

\section{The effective Hamiltonian of this system~\label{app:A}}

We input signals, which include both the TE and TM modes, and suppose  the coupling strength between TM mode and mechanical mode($g_{a}$) far less than that of TE mode and mechanical mode($g_{b}$), i.e. $g_{a}\ll g_{b}$. In this sense, the Hamiltonian is,
\begin{equation}
H=\omega_{a} \hat{a}^{\dag}\hat{a}+\omega_{b}\hat{b}^{\dag}\hat{b}+\omega_{m}\hat{m}^{\dag}\hat{m}
+\omega_{r}\hat{r}^{\dag}\hat{r}+g_{m}(\hat{a}\hat{b}^{\dag}\hat{m}+\hat{a}^{\dag}\hat{b}\hat{m}^{\dag})+g_{b}\hat{b}^{\dag}\hat{b}(\hat{r}^{\dag}+\hat{r}),
\end{equation}
where $\omega_{a}$($\omega_{b}$) is the frequency of the TM(TE) mode, $\omega_{m}$ is the magnetic frequency and $\omega_{r}$ is the frequency of the phonon. The $j^{\dag}$($j$) $(j=a,b,m,r)$ is the creation (annihilation) operator of corresponding mode(TM mode photon, TE mode photon, magnon and phonon). $g_{m}$ is the interaction strength of  photon and magnon, and $g_{b}$ is the interaction strength of photon and phonon. The TM  mode is driven by $\sqrt{2\kappa_{a,1}}a_{in}(\hat{a}e^{i \omega_{L}t}+\hat{a}^{\dag}e^{-i\omega_{L}t})$.  $\kappa_{a,1}$ is the coupling damping. $a_{in}$ is the input power and $\omega_{L}$ is the TM pumping frequency. With the non-depletion approximation, the intracavity TM field
\begin{equation}
\langle a\rangle=\frac{\sqrt{2 \kappa_{a,1}}a_{in}}{-i(\omega_{L}-\omega_{a})+\kappa_{a}},
\end{equation}
where the $\kappa_{a}$ is the total damping of the TM mode of the cavity. The $\langle a\rangle$ corresponds to the interaction enhancement of the interaction between the TE-photon and magnon mode. We write this effective coupling strength as $G_{a}=g_{m}\langle a\rangle$. Then the effective Hamiltonian can be read as
\begin{equation}
H=\omega_{b}\hat{b}^{\dag}\hat{b}+\omega_{m}\hat{m}^{\dag}\hat{m}+\omega_{r}\hat{r}^{\dag}\hat{r}+G_{a}\hat{b}^{\dag}\hat{m}+G_{a}^{*}\hat{b} \hat{m}^{\dag}+g\hat{b}^{\dag}\hat{b}(
\hat{r}^{\dag}+\hat{r}).
\end{equation}
Linear the optomechanical interaction under the strong red-side band pumping $\sqrt{2\kappa_{b,1}}b_{in}(\hat{b}e^{i \omega_{pu}t}+\hat{b}^{\dag}e^{-i\omega_{pu}t})$. We rotate the system with the steady TE field $\bar{b}$. When omit the term which shift the mechanical resonator’s equilibrium position and high order perturbation, the effective Hamiltonian have the form
\begin{equation}
H=\omega_{b}\hat{b}^{\dag}\hat{b}+\omega_{m}\hat{m}^{\dag}\hat{m}+\omega_{r}\hat{r}^{\dag}\hat{r}+G_{a}\hat{b}^{\dag}\hat{m}+G_{a}^{*}\hat{b} \hat{m}^{\dag}+G_{b}(\hat{b}^{\dag}+\hat{b})\hat{z},
\end{equation}
where $\hat{z}=\hat{r}^{\dag}+\hat{r}$. The effective coupling strength $G_{b}=\bar{b}g_{b}=\sqrt{2\kappa_{b,1}}b_{in}/(\kappa-i(\omega_{pu}-\omega_{b}))$. In this Hamiltonian $\hat{b}$ is the perturbation term of the TE field.
\section{The coupling strength in the frequency domain~\label{App;B}}

In common sense, the room-temperature experiment is firmly in the classical regime $\hbar\omega_{a}\gg k_{B}T$, we use a classical analysis. We also disregard all noise on the optical drive, applying the rotating frame of $\omega_{pu}\hat{b}^{\dag}\hat{b}$, we use $\Delta_{r}=\omega_{pu}-\omega_{b}$ for simplicity. Then the kinetic equations of this system can be expressed as
\begin{subequations}
	\begin{align}
	\dot{b}&=-(\frac{\kappa}{2}-i\Delta_{b})b-iG_{b}z-iG_{a}m ,\\
	\dot{r}&=-(\frac{\gamma_{r}}{2}+i\omega_{r})r-i(G_{b}^{*}b+G_{b}b^{*})+\sqrt{\gamma_{r}}\eta_{r}, \\
	\dot{m}&=-(\frac{\gamma_{m}}{2}+i\omega_{m})r-iG_{a}b+\sqrt{\gamma_{m}}\eta_{m}.
	\end{align}
\end{subequations}
The equation in the frequency domain can be derived with the Fourier transform. Here the magnon and phonon is bathed with thermal fluctuations, the  thermal Langevin force $\eta_{i}(t)$ is described by $\langle\eta_{i}(t)\rangle=0$ and $\langle\eta_{j}^{*}(t)\eta_{i}(t)\rangle=\delta_{ij}\delta{t-t'}k_{B}T/\hbar\omega_{i}$
\begin{subequations}
	\begin{align}
	\chi_{b}^{-1}[\omega]b[\omega]&=-iG_{b}z[\omega]-iG_{a}m[\omega],\\
	\chi_{r}^{-1}[\omega]r[\omega]&=-i(G_{b}^{*}b[\omega]+G_{b}b^{*}[-\omega])+\sqrt{\gamma_{r}}\eta_{r}[\omega],\\
	\chi_{m}^{-1}[\omega]m[\omega]&=-iG_{a}b[\omega]+\sqrt{\gamma_{m}}\eta_{m}[\omega].
	\end{align}
	\end{subequations}
The $\chi[\omega]=(\gamma/2-i(\omega-\omega_{j})),j=b,r,m$  the intrinsic susceptibility of its corresponding degree of freedom.

\section{The reduced Hamiltonian~\label{App;C}}

 With rotating wave approximation for $\omega_{r}$ and $\omega_{m}$. Eliminate the electromagnetic freedom. Eq.\ref{effective} in the main text can be written as,
 \begin{subequations}
 	\begin{align}
 \chi_{r}^{-1}[\omega]r[\omega]=&-|G_{b}|^{2}[\chi_{b}[\omega]-\chi_{b}^{*}[-\omega]]r[\omega]-G_{a}^{*}G_{b}\chi_{b}[\omega]m[\omega],
 \\
 	\chi_{m}^{-1}[\omega]m[\omega]=&-G_{a}^{2}\chi_{b}[\omega]r[\omega]-G_{a}G_{b}\chi_{b}[\omega]r[\omega].
 	\end{align}
 \end{subequations}
 As a closed system, we discard the thermal noise term. And it can be achieved by putting this system in the low temperature environment. In this two mode system we define the self-energy as
 \begin{equation}
 \mathbf{\Sigma}[\omega]=-i\left\{\begin{matrix} |G_{b}|^{2}[\chi_{b}[\omega]-\chi_{b}^{*}[-\omega]] & G_{a}^{*}G_{b}\chi_{b}[\omega]  \\ G_{a}G_{b}\chi_{b}[\omega] & G_{a}^{2}\chi_{b}[\omega] \end{matrix}\right\}.
 \end{equation}

 When combine the phonon and magnon as a whole vector $\mathbf{s}(t)=(r(t),m(t))^{T}$. Then the system can be written in the matrix form,
 \begin{equation}
 -i\omega\mathbf{s}[\omega]=-\left\{\begin{matrix} \frac{\kappa_{r}}{2}+i\omega_{r} & 0  \\ 0 &  \frac{\kappa_{m}}{2}+i\omega_{m} \end{matrix}\right\}\mathbf{s}[\omega]-i\mathbf{\Sigma}[\omega]\mathbf{s}[\omega].
 \end{equation}
 It is noticeable that $\mathbf{\Sigma}[\omega]$ various on the scale of $\gamma$. whereas this vector modes are susceptible to drives only if their
 is substantially smaller than $\gamma$.  Therefore, it is sufficient to consider  $\mathbf{\Sigma}[\omega]\approx \mathbf{\Sigma}[\omega_{r}] \approx\mathbf{\Sigma}[\omega_{m}]\equiv\mathbf{\Sigma}$. Then self-energy is frequency independence. So we can easily move back to the time domain to get the effective schr\H{o}dinger's equation,
 \begin{equation}
 i\dot{\mathbf{s}}(t)=H\mathbf{s}(t).
 \end{equation}
 Here the Hamiltonian reads,
 \begin{equation}
 H=\left\{\begin{matrix} \omega_{r}-i\frac{ \kappa_{r}}{2} & 0  \\ 0 & \omega_{m}-i \frac{ \kappa_{m}}{2} \end{matrix}\right\}+\mathbf{\Sigma}.
 \end{equation}

\section*{References}
\bibliographystyle{iopart-num}
\bibliography{ref}

\providecommand{\newblock}{}
\begin{thebibliography}{10}
\expandafter\ifx\csname url\endcsname\relax
  \def\url#1{{\tt #1}}\fi
\expandafter\ifx\csname urlprefix\endcsname\relax\def\urlprefix{URL }\fi
\providecommand{\eprint}[2][]{\url{#2}}

\bibitem{Bowen-Milburn}
Warwick P and Gerard J 2016 {\em Quantum optomechanics\/} (CRC Press)
  \urlprefix\url{https://www.crcpress.com/Quantum-Optomechanics/Bowen-Milburn/p/book/9781482259155}

\bibitem{RevModPhys.86.1391}
Aspelmeyer M, Kippenberg T and Marquardt F 2014 {\em Rev. Mod. Phys.\/} {\bf
  86}(4) 1391--1452
  \urlprefix\url{http://link.aps.org/doi/10.1103/RevModPhys.86.1391}

\bibitem{Aspelmeyer}
Aspelmeyer M, Kippenberg T and Marquardt F 2014 {\em Cavity Optomechanics:
  Nano- and Micromechanical Resonators Interacting with Light\/} 1st ed Quantum
  Science and Technology (Springer-Verlag Berlin Heidelberg)
  \urlprefix\url{http://www.springer.com/gp/book/9783642553110}

\bibitem{Kippenberg:07}
Kippenberg T and Vahala K 2007 {\em Opt. Express\/} {\bf 15} 17172--17205
  \urlprefix\url{http://www.opticsexpress.org/abstract.cfm?URI=oe-15-25-17172}

\bibitem{peng2014and}
Peng B, {\"O}zdemir {\c{S}}~K, Chen W, Nori F and Yang L 2014 {\em Nature
  communications\/} {\bf 5} 5082
  \urlprefix\url{http://www.nature.com/articles/ncomms6082}

\bibitem{anetsberger2009near}
Anetsberger G, Arcizet O, Unterreithmeier Q~P, Rivi{\`e}re R, Schliesser A,
  Weig E~M, Kotthaus J~P and Kippenberg T~J 2009 {\em Nature Physics\/} {\bf 5}
  909--914
  \urlprefix\url{http://www.nature.com/nphys/journal/v5/n12/abs/nphys1425.html}

\bibitem{Kippenberg1172}
Kippenberg T and Vahala K 2008 {\em Science\/} {\bf 321} 1172--1176
  \urlprefix\url{http://science.sciencemag.org/content/321/5893/1172}

\bibitem{hill2012coherent}
Hill J~T, Safavi-Naeini A~H, Chan J and Painter O 2012 {\em Nature
  communications\/} {\bf 3} 1196
  \urlprefix\url{http://www.nature.com/articles/ncomms2201}

\bibitem{PhysRevLett.110.153606}
Liu Y~C, Xiao Y~F, Luan X and Wong C~W 2013 {\em Phys. Rev. Lett.\/} {\bf
  110}(15) 153606
  \urlprefix\url{http://link.aps.org/doi/10.1103/PhysRevLett.110.153606}

\bibitem{PhysRevLett.98.150802}
Corbitt T, Chen Y, Innerhofer E, M\"uller-Ebhardt H, Ottaway D, Rehbein H, Sigg
  D, Whitcomb S, Wipf C and Mavalvala N 2007 {\em Phys. Rev. Lett.\/} {\bf
  98}(15) 150802
  \urlprefix\url{http://link.aps.org/doi/10.1103/PhysRevLett.98.150802}

\bibitem{thompson2008strong}
Thompson J, Zwickl B, Jayich A, Marquardt F, Girvin S and Harris J 2008 {\em
  Nature\/} {\bf 452} 72--75
  \urlprefix\url{http://www.nature.com/nature/journal/v452/n7183/abs/nature06715.html}

\bibitem{PhysRevLett.113.156401}
Zhang X, Zou C~L, Jiang L and Tang H~X 2014 {\em Phys. Rev. Lett.\/} {\bf
  113}(15) 156401
  \urlprefix\url{http://link.aps.org/doi/10.1103/PhysRevLett.113.156401}

\bibitem{PhysRevB.94.060405}
Liu T, Zhang X, Tang H~X and Flatt\'e M~E 2016 {\em Phys. Rev. B\/} {\bf 94}(6)
  060405 \urlprefix\url{http://link.aps.org/doi/10.1103/PhysRevB.94.060405}

\bibitem{PhysRevLett.117.123605}
Zhang X, Zhu N, Zou C~L and Tang H~X 2016 {\em Phys. Rev. Lett.\/} {\bf
  117}(12) 123605
  \urlprefix\url{http://link.aps.org/doi/10.1103/PhysRevLett.117.123605}

\bibitem{PhysRevA.92.063845}
Haigh J, Langenfeld S, Lambert N, Baumberg J, Ramsay A, Nunnenkamp A and
  Ferguson A 2015 {\em Phys. Rev. A\/} {\bf 92}(6) 063845
  \urlprefix\url{http://link.aps.org/doi/10.1103/PhysRevA.92.063845}

\bibitem{PhysRevB.93.144420}
Bourhill J, Kostylev N, Goryachev M, Creedon D and Tobar M 2016 {\em Phys. Rev.
  B\/} {\bf 93}(14) 144420
  \urlprefix\url{http://link.aps.org/doi/10.1103/PhysRevB.93.144420}

\bibitem{PhysRevLett.116.223601}
Osada A, Hisatomi R, Noguchi A, Tabuchi Y, Yamazaki R, Usami K, Sadgrove M,
  Yalla R, Nomura M and Nakamura Y 2016 {\em Phys. Rev. Lett.\/} {\bf 116}(22)
  223601 \urlprefix\url{http://link.aps.org/doi/10.1103/PhysRevLett.116.223601}

\bibitem{PhysRevLett.117.133602}
Haigh J, Nunnenkamp A, Ramsay A and Ferguson A 2016 {\em Phys. Rev. Lett.\/}
  {\bf 117}(13) 133602
  \urlprefix\url{http://link.aps.org/doi/10.1103/PhysRevLett.117.133602}

\bibitem{PhysRevA.94.033821}
Viola~Kusminskiy S, Tang H~X and Marquardt F 2016 {\em Phys. Rev. A\/} {\bf
  94}(3) 033821
  \urlprefix\url{http://link.aps.org/doi/10.1103/PhysRevA.94.033821}

\bibitem{klos2014photonic}
K{\l}os J~W, Krawczyk M, Dadoenkova Y~S, Dadoenkova N and Lyubchanskii I 2014
  {\em Journal of Applied Physics\/} {\bf 115} 174311
  \urlprefix\url{"http://scitation.aip.org/content/aip/journal/jap/115/17/10.1063/1.4874797"}

\bibitem{chumak2015magnon}
Chumak A, Vasyuchka V, Serga A and Hillebrands B 2015 {\em Nature Physics\/}
  {\bf 11} 453--461
  \urlprefix\url{http://www.nature.com/nphys/journal/v11/n6/abs/nphys3347.html}

\bibitem{Tabuchi405}
Tabuchi Y, Ishino S, Noguchi A, Ishikawa T, Yamazaki R, Usami K and Nakamura Y
  2015 {\em Science\/} {\bf 349} 405--408
  \urlprefix\url{http://science.sciencemag.org/content/349/6246/405}

\bibitem{PhysRevLett.111.127003}
Huebl H, Zollitsch C~W, Lotze J, Hocke F, Greifenstein M, Marx A, Gross R and
  Goennenwein S~T 2013 {\em Phys. Rev. Lett.\/} {\bf 111}(12) 127003
  \urlprefix\url{http://link.aps.org/doi/10.1103/PhysRevLett.111.127003}

\bibitem{brooks2012non}
Brooks D~W, Botter T, Schreppler S, Purdy T~P, Brahms N and Stamper-Kurn D~M
  2012 {\em Nature\/} {\bf 488} 476--480
  \urlprefix\url{http://www.nature.com/nature/journal/v488/n7412/abs/nature11325.html}

\bibitem{safavi2013squeezed}
Safavi-Naeini A~H, Gr{\"o}blacher S, Hill J~T, Chan J, Aspelmeyer M and Painter
  O 2013 {\em Nature\/} {\bf 500} 185--189
  \urlprefix\url{http://www.nature.com/nature/journal/v500/n7461/abs/nature12307.html}

\bibitem{PhysRevX.3.031012}
Purdy T~P, Yu P~L, Peterson R~W, Kampel N~S and Regal C~A 2013 {\em Phys. Rev.
  X\/} {\bf 3}(3) 031012
  \urlprefix\url{http://link.aps.org/doi/10.1103/PhysRevX.3.031012}

\bibitem{PhysRevA.83.033820}
Liao J~Q and Law C~K 2011 {\em Phys. Rev. A\/} {\bf 83}(3) 033820
  \urlprefix\url{http://link.aps.org/doi/10.1103/PhysRevA.83.033820}

\bibitem{PhysRevA.82.021806}
Nunnenkamp A, B\o{}rkje K, Harris J~G~E and Girvin S~M 2010 {\em Phys. Rev.
  A\/} {\bf 82}(2) 021806
  \urlprefix\url{http://link.aps.org/doi/10.1103/PhysRevA.82.021806}

\bibitem{PhysRevLett.97.133601}
Arcizet O, Cohadon P~F, Briant T, Pinard M, Heidmann A, Mackowski J~M, Michel
  C, Pinard L, Fran\ifmmode~\mbox{\c{c}}\else \c{c}\fi{}ais O and Rousseau L
  2006 {\em Phys. Rev. Lett.\/} {\bf 97}(13) 133601
  \urlprefix\url{http://link.aps.org/doi/10.1103/PhysRevLett.97.133601}

\bibitem{PhysRevLett.105.123601}
Tsang M and Caves C~M 2010 {\em Phys. Rev. Lett.\/} {\bf 105}(12) 123601
  \urlprefix\url{http://link.aps.org/doi/10.1103/PhysRevLett.105.123601}

\bibitem{teufel2009nanomechanical}
Teufel J, Donner T, Castellanos-Beltran M, Harlow J and Lehnert K 2009 {\em
  Nature nanotechnology\/} {\bf 4} 820--823
  \urlprefix\url{http://www.nature.com/nnano/journal/v4/n12/abs/nnano.2009.343.html}

\bibitem{PhysRevLett.113.020405}
Nimmrichter S, Hornberger K and Hammerer K 2014 {\em Phys. Rev. Lett.\/} {\bf
  113}(2) 020405
  \urlprefix\url{http://link.aps.org/doi/10.1103/PhysRevLett.113.020405}

\bibitem{PhysRevLett.112.013602}
Shkarin A~B, Flowers-Jacobs N~E, Hoch S~W, Kashkanova A~D, Deutsch C, Reichel J
  and Harris J~G~E 2014 {\em Phys. Rev. Lett.\/} {\bf 112}(1) 013602
  \urlprefix\url{http://link.aps.org/doi/10.1103/PhysRevLett.112.013602}

\bibitem{Weis1520}
Weis S, Rivi{\`e}re R, Del{\'e}glise S, Gavartin E, Arcizet O, Schliesser A and
  Kippenberg T~J 2010 {\em Science\/} {\bf 330} 1520--1523
  \urlprefix\url{http://science.sciencemag.org/content/330/6010/1520}

\bibitem{monifi2016optomechanically}
Monifi F, Zhang J, {\"O}zdemir {\c{S}}~K, Peng B, Liu Y~x, Bo F, Nori F and
  Yang L 2016 {\em Nature Photonics\/}  399–--405
  \urlprefix\url{http://www.nature.com/nphoton/journal/v10/n6/full/nphoton.2016.73.html}

\bibitem{safavi2011electromagnetically}
Safavi-Naeini A~H, Alegre T~M, Chan J, Eichenfield M, Winger M, Lin Q, Hill
  J~T, Chang D~E and Painter O 2011 {\em Nature\/} {\bf 472} 69--73
  \urlprefix\url{http://www.nature.com/nature/journal/v472/n7341/abs/nature09933.html}

\bibitem{jing2015optomechanically}
Jing H, {\"O}zdemir {\c{S}}~K, Geng Z, Zhang J, L{\"u} X~Y, Peng B, Yang L and
  Nori F 2015 {\em Scientific reports\/} {\bf 5} 9663
  \urlprefix\url{http://www.nature.com/articles/srep09663}

\bibitem{PhysRevLett.109.013603}
Stannigel K, Komar P, Habraken S~J~M, Bennett S~D, Lukin M~D, Zoller P and Rabl
  P 2012 {\em Phys. Rev. Lett.\/} {\bf 109}(1) 013603
  \urlprefix\url{http://link.aps.org/doi/10.1103/PhysRevLett.109.013603}

\bibitem{Ying-Dan}
Wang Y~D and Clerk A~A 2012 {\em New Journal of Physics\/} {\bf 14} 105010
  \urlprefix\url{http://stacks.iop.org/1367-2630/14/i=10/a=105010}

\bibitem{PhysRevLett.111.083601}
Liu Y~C, Xiao Y~F, Chen Y~L, Yu X~C and Gong Q 2013 {\em Phys. Rev. Lett.\/}
  {\bf 111}(8) 083601
  \urlprefix\url{http://link.aps.org/doi/10.1103/PhysRevLett.111.083601}

\bibitem{PhysRevLett.109.063601}
Ludwig M, Safavi-Naeini A~H, Painter O and Marquardt F 2012 {\em Phys. Rev.
  Lett.\/} {\bf 109}(6) 063601
  \urlprefix\url{http://link.aps.org/doi/10.1103/PhysRevLett.109.063601}

\bibitem{Serga}
Serga A~A, Chumak A~V and Hillebrands B 2010 {\em Journal of Physics D: Applied
  Physics\/} {\bf 43} 264002
  \urlprefix\url{http://stacks.iop.org/0022-3727/43/i=26/a=264002}

\bibitem{PhysRevLett.110.147206}
Montagnese M, Otter M, Zotos X, Fishman D~A, Hlubek N, Mityashkin O, Hess C,
  Saint-Martin R, Singh S, Revcolevschi A and van Loosdrecht P~H~M 2013 {\em
  Phys. Rev. Lett.\/} {\bf 110}(14) 147206
  \urlprefix\url{http://link.aps.org/doi/10.1103/PhysRevLett.110.147206}

\bibitem{lenk2011building}
Lenk B, Ulrichs H, Garbs F and M{\"u}nzenberg M 2011 {\em Physics Reports\/}
  {\bf 507} 107--136
  \urlprefix\url{http://www.sciencedirect.com/science/article/pii/S0370157311001694}

\bibitem{Cowburn1466}
Cowburn R~P and Welland M~E 2000 {\em Science\/} {\bf 287} 1466--1468
  \urlprefix\url{http://science.sciencemag.org/content/287/5457/1466}

\bibitem{Imre205}
Imre A, Csaba G, Ji L, Orlov A, Bernstein G~H and Porod W 2006 {\em Science\/}
  {\bf 311} 205--208
  \urlprefix\url{http://science.sciencemag.org/content/311/5758/205}

\bibitem{PhysRevLett.103.043601}
Tanji H, Ghosh S, Simon J, Bloom B and Vuleti\ifmmode~\acute{c}\else \'{c}\fi{}
  V 2009 {\em Phys. Rev. Lett.\/} {\bf 103}(4) 043601
  \urlprefix\url{http://link.aps.org/doi/10.1103/PhysRevLett.103.043601}

\bibitem{PhysRevLett.74.1867}
Lorenzana J and Sawatzky G~A 1995 {\em Phys. Rev. Lett.\/} {\bf 74}(10)
  1867--1870
  \urlprefix\url{http://link.aps.org/doi/10.1103/PhysRevLett.74.1867}

\bibitem{zhang2015magnon}
Zhang X, Zou C~L, Zhu N, Marquardt F, Jiang L and Tang H~X 2015 {\em Nature
  communications\/} {\bf 6} 8914
  \urlprefix\url{http://www.nature.com/articles/ncomms9914}

\bibitem{armani2003ultra}
Armani D, Kippenberg T, Spillane S and Vahala K 2003 {\em Nature\/} {\bf 421}
  925--928
  \urlprefix\url{http://www.nature.com/nature/journal/v421/n6926/abs/nature01371.html}

\bibitem{vahala2003optical}
Vahala K~J 2003 {\em Nature\/} {\bf 424} 839--846
  \urlprefix\url{http://www.nature.com/nature/journal/v424/n6950/abs/nature01939.html}

\bibitem{gorodetsky1999optical}
Gorodetsky M~L and Ilchenko V~S 1999 {\em JOSA B\/} {\bf 16} 147--154
  \urlprefix\url{https://www.osapublishing.org/josab/abstract.cfm?uri=josab-16-1-147}

\bibitem{Magneto-Optics}
Kushida T 2000 {\em Magneto-Optics\/} 1st ed Springer Series in Solid-State
  Sciences 128 (Springer-Verlag Berlin Heidelberg)

\bibitem{PhysRev.137.A448}
Bennett H~S and Stern E~A 1965 {\em Phys. Rev.\/} {\bf 137}(2A) A448--A461
  \urlprefix\url{http://link.aps.org/doi/10.1103/PhysRev.137.A448}

\bibitem{inoue1997theoretical}
Inoue M and Fujii T 1997 {\em Journal of applied physics\/} {\bf 81} 5659--5661
  \urlprefix\url{http://scitation.aip.org/content/aip/journal/jap/81/8/10.1063/1.364687}

\bibitem{RevModPhys.82.2731}
Kirilyuk A, Kimel A~V and Rasing T 2010 {\em Rev. Mod. Phys.\/} {\bf 82}(3)
  2731--2784 \urlprefix\url{http://link.aps.org/doi/10.1103/RevModPhys.82.2731}

\bibitem{kimel2005ultrafast}
Kimel A, Kirilyuk A, Usachev P, Pisarev R, Balbashov A and Rasing T 2005 {\em
  Nature\/} {\bf 435} 655--657
  \urlprefix\url{http://www.nature.com/nature/journal/v435/n7042/abs/nature03564.html}

\bibitem{Demokritov2001441}
Demokritov S, Hillebrands B and Slavin A 2001 {\em Physics Reports\/} {\bf 348}
  441 -- 489
  \urlprefix\url{http://www.sciencedirect.com/science/article/pii/S0370157300001162}

\bibitem{PhysRev.166.514}
Fleury P~A and Loudon R 1968 {\em Phys. Rev.\/} {\bf 166}(2) 514--530
  \urlprefix\url{http://link.aps.org/doi/10.1103/PhysRev.166.514}

\bibitem{GORODETSKY1994133}
Gorodetsky M and Ilchenko V 1994 {\em Optics Communications\/} {\bf 113} 133 --
  143
  \urlprefix\url{http://www.sciencedirect.com/science/article/pii/0030401894906033}

\end{thebibliography}

\end{document}